# Conductance saturation in a series of highly transmitting molecular junctions

T. Yelin[1], R. Korytár[2], N. Sukenik[1], R. Vardimon[1], B. Kumar[3], C. Nuckolls[3], F. Evers[2], O. Tal[1]

1. Department of Chemical Physics, Weizmann Institute of Science, Rehovot, Israel.
2. Institut für Theoretische Physik, Universität Regensburg, Regensburg, Germany.
3. Department of Chemistry, Columbia University, New York, United States.

**Note:** The published version in Nature Materials contains an extended manuscript and comprehensive supplementary information.

**Understanding the properties of electronic transport across metal-molecule interfaces is of central importance for controlling a large variety of molecular-based devices such as organic light emitting diodes[1], nanoscale organic spin-valves[2] and single-molecule switches[3]. One of the primary experimental methods to reveal the mechanisms behind electronic transport through metal-molecule interfaces is the study of conductance as a function of molecule length in molecular junctions[4-14]. Previous studies focused on transport governed either by coherent tunneling or hopping, both at low conductance. However, the upper limit of conductance across molecular junctions has not been explored, despite the great potential for efficient information transfer, charge injection and recombination processes at high conductance. Here, we study the conductance properties of highly transmitting metal-molecule-metal interfaces, using a series of single-molecule junctions based on oligoacenes with increasing length. We find that the conductance saturates at an upper limit where it is independent of molecule length. Furthermore, we show that this upper limit can be controlled by the character of the orbital hybridization at the metal-molecule interface. Using two prototype systems, in which the molecules are contacted by either Ag or Pt electrodes, we reveal two different origins for the saturation of conductance. In the case of Ag-based molecular junctions, the conductance saturation is ascribed to a competition between energy level alignment and level broadening, while in the case of Pt-based junctions, the saturation is attributed to a band-like transport. The results are explained by an intuitive model, backed by *ab-initio* transport calculations. Our findings shed light on the mechanisms that constrain the conductance at the high transmission limit, providing guiding principles for the design of highly conductive metal-molecule interfaces.**

In order to study the conductance characteristics of highly transmitting molecular junctions, strong electronic coupling is required between the molecule and the electrodes, as well as within the molecule itself[10, 11, 15, 16]. These conditions are achieved in this work by direct hybridization between the π-orbitals of the oligoacene molecules and the frontier orbitals of the metal electrodes, without employing anchoring groups such as thiols that can act as spacers between the orbitals of the molecular backbone and the frontier orbitals of the metal[13, 15]. The oligoacenes (Fig. 1a) are linear π-conjugated molecules that can be viewed as short graphene nanoribbons[17], whose electronic structure is subject to an ongoing research[18]. We study the evolution of conductance as a function of molecule length and compare the conductance characteristics of two prototype systems based on either Ag or Pt electrodes. While Ag has frontier *s*-orbitals available for conductance, Pt has also prominent frontier *d*-orbitals. Ag/oligoacene junctions yield a non-trivial conductance trend, where the conductance first increases with molecule length, followed by the onset of conductance saturation. Conversely, the conductance of Pt/oligoacene junctions does not depend on the molecule length, and is approximately equal to the conductance quantum ($G_0$=2 $e^2/h$≈ (12.9 kΩ)$^{-1}$). Our analysis indicates that Ag-based molecular junctions preserve the fingerprint of the electronic structure of the molecules and serve as a probe for the interplay between level alignment and coupling strength. In contrast, for Pt-based junctions, the molecular energetic features are smeared out, yielding a band-like transmission. These findings pave the way for controlling electronic transport at the high transmission limit, by manipulating the orbital hybridization at metal-molecule interfaces.

Experiments are performed in a break-junction set-up[19] (Fig. 1a) in cryogenic temperature (~4.2K). Two freshly exposed electrodes are formed by controllably breaking a metal wire into two segments, separated by a gap that can be adjusted in sub-atomic resolution. Before introducing the molecules, the typical conductance of the metal atomic junction is analyzed by recording conductance traces as a function of the relative electrode displacement, as shown for Ag junction in Fig. 1b, left panel. When the electrodes are pulled apart, the contact is narrowed, and the conductance decreases ("pull" conductance traces, blue). The conductance plateaus at ~1 $G_0$ correspond to a single Ag atom at the smallest cross-section[20, 21], while further stretching leads to junction rupture. Following the rupture of the junction, the electrodes are pushed together and the conductance increases as the contact is reformed ("push" conductance traces, red). To characterize the key conductance features, conductance histograms are constructed from thousands of traces. Fig. 1c, left panel, reveals peaks at ~1 $G_0$ that correspond to the most probable conductance value of the atomic Ag junction. The tail at low conductance is the signature of tunneling transport measured right after rupture or before reforming an atomic contact[21, 22]. The density plots in Fig. 1c, middle and right panels, show the conductance evolution when pulling or pushing the electrodes, respectively.

In a series of independent experiments, oligoacene molecules (benzene, naphthalene, anthracene…) are introduced, one type of oligoacenes at each experiment, to a bare Ag junction from a molecular source (Fig. 1a). As an example, the conductance characterization of Ag/anthracene single-molecule junctions is shown in Fig. 1d. The peak at ~1 $G_0$ is still apparent, however, an additional peak is observed at lower conductance, indicating the formation of molecular junctions. Typical conductance traces are shown in Fig. 1b, right panel. The characteristic conductance of each Ag/oligoacene molecular junction, determined from the peaks in the conductance histograms, is presented in Fig. 2. Out of the six molecules studied, five showed a clear peak in the conductance histograms. The conductance trend along the Ag/oligoacene series is composed of two main regions. Initially, the conductance increases with molecule length. However, beyond three rings, the onset of conductance saturation can be observed. While conductance increase with molecule length was previously reported[7, 11], this is the first observation of conductance saturation.

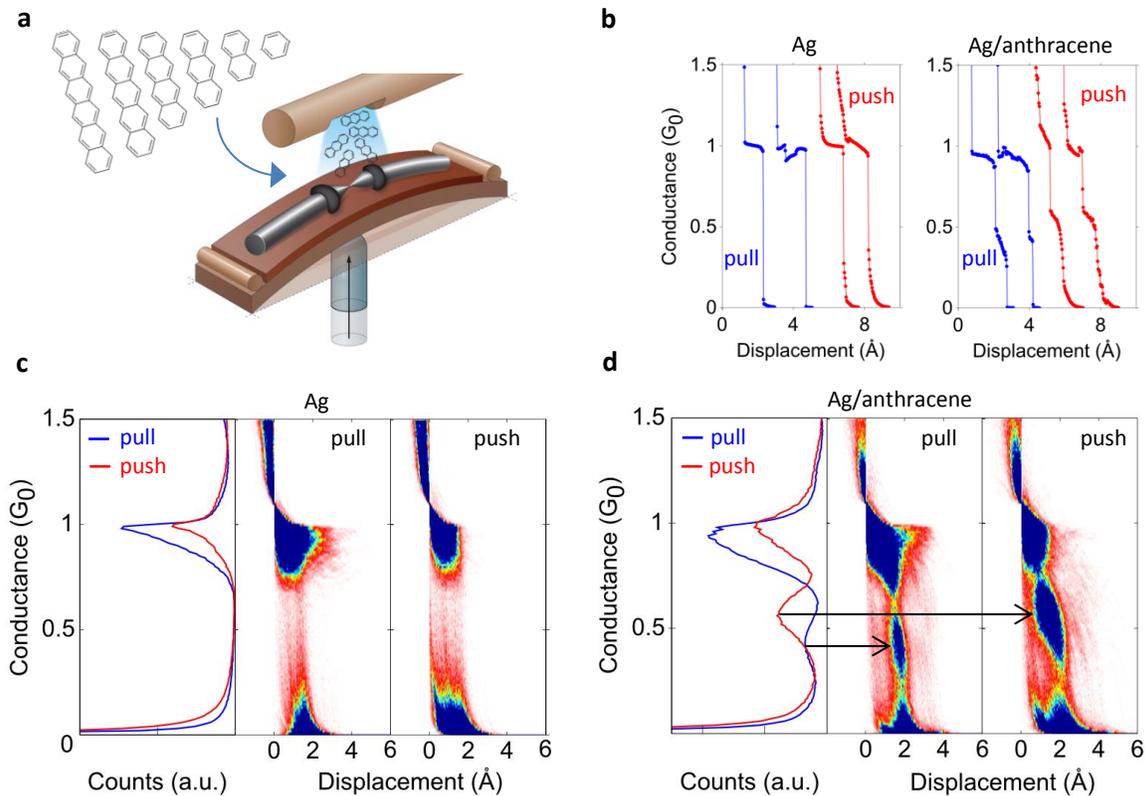

**Figure 1. Characterization of Ag/oligoacene molecular junctions**. **a.** Illustration of a break junction setup and structure of the studied series of oligoacene molecules (right to left: benzene, naphthalene, anthracene, tetracene, pentacene, hexacene) **b.** Pull (blue) and push (red) conductance traces of Ag (left panel) and Ag/anthracene (right panel). **c.** Characterization of Ag atomic junctions. Left panel: pull (blue) and push (red) conductance histograms. Middle and right panels: conductance-displacement density plots constructed from pull and push conductance traces, respectively. The histograms and the density plots are composed from at least 5,000 traces, recorded at a bias voltage of 100 mV. Zero displacement is set for each trace as the first displacement point with a conductance value below 1.1 $G_0$. **d.** Same for Ag/anthracene molecular junctions. Push values are typically higher than pull values, probably due to induced elongation of the bonds in the pulling process that suppresses the conductance.

To understand the origin of the conductance trend along the Ag/oligoacene series, we performed *Ab-initio* transport calculations. The calculations show that due to the direct coupling, the molecule has the freedom to align with the electrodes in a variety of configurations. Fig. 3a presents the transmission curves for a compact configuration of the molecule (Fig. 3a, inset), with its long axis pointing out of the electrodes axis. This configuration yields the best fit to the experimental values along the Ag/oligoacene series. Moreover, the measured total elongation of the molecular junctions shows no correlation with molecule length, consistent with a compact configuration, for which the length of the molecule is not manifested in the junction length. The conductance, determined by the value of the transmission curves at the Fermi energy $E_F$ (see Fig.3a), is dominated by the lowest unoccupied molecular orbitals (LUMO) which are much closer to $E_F$ than the highest occupied molecular orbitals (HOMO). The difference between the transmission values at $E_F$ diminishes with molecule length, implying the onset of conductance saturation. The transmission peaks can be approximated by Lorentzians, except for the benzene peak, due to a two-fold LUMO degeneracy. The evolution of the LUMO peaks with molecule length points on two trends with competing effects on the conductance. On one hand, the reduction in the separation between $E_F$ and the peak center $\varepsilon$ (Fig. 3b, red) acts to increase the conductance, on the other hand, the reduction in the peak width $\Gamma$ (Fig. 3b, blue) acts to decrease the conductance.

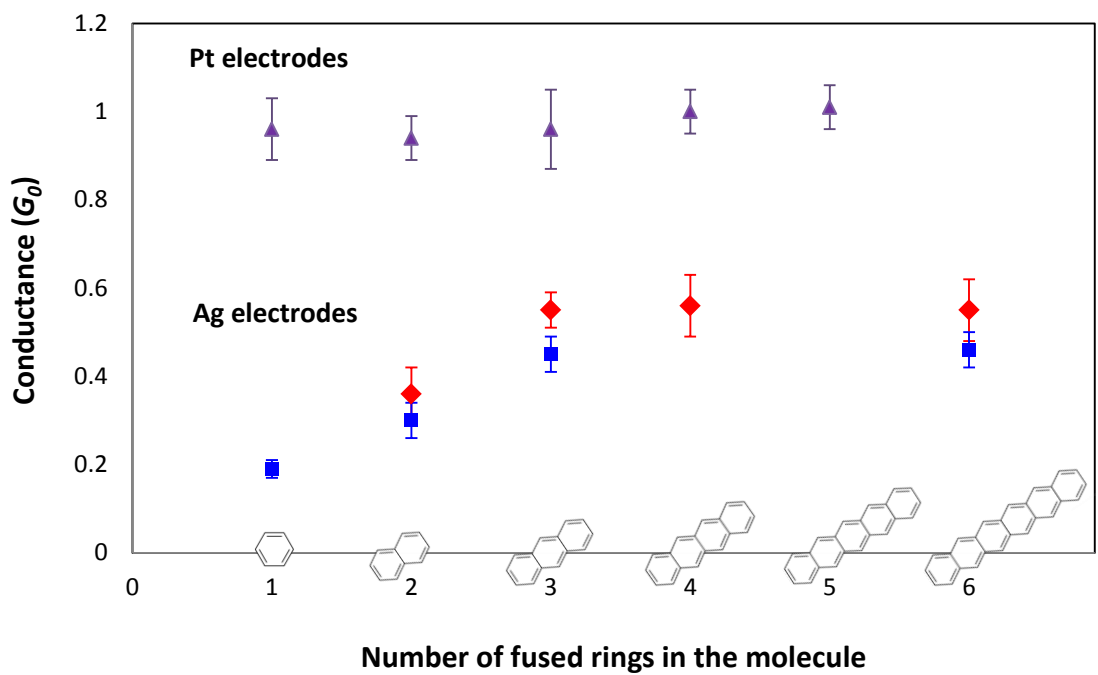

**Figure 2: The characteristic conductance of Ag/oligoacene and Pt/oligoacene junctions as a function of the number of benzene rings in the oligoacene molecule. Lower series:** Most probable conductance of Ag/oligoacenes junctions, determined from the peaks in the pull (■) and push (♦) conductance histograms. All the studied molecules except the pentacene showed a clear conductance peak in either pull or push directions. **Upper series:** Most probable conductance of Pt/oligoacenes junctions, determined from the pull conductance histograms (▲). Values from push histograms (not presented) are typically higher by ~0.02-0.1 $G_0$. Error bars represent the standard deviation between independent experiments.

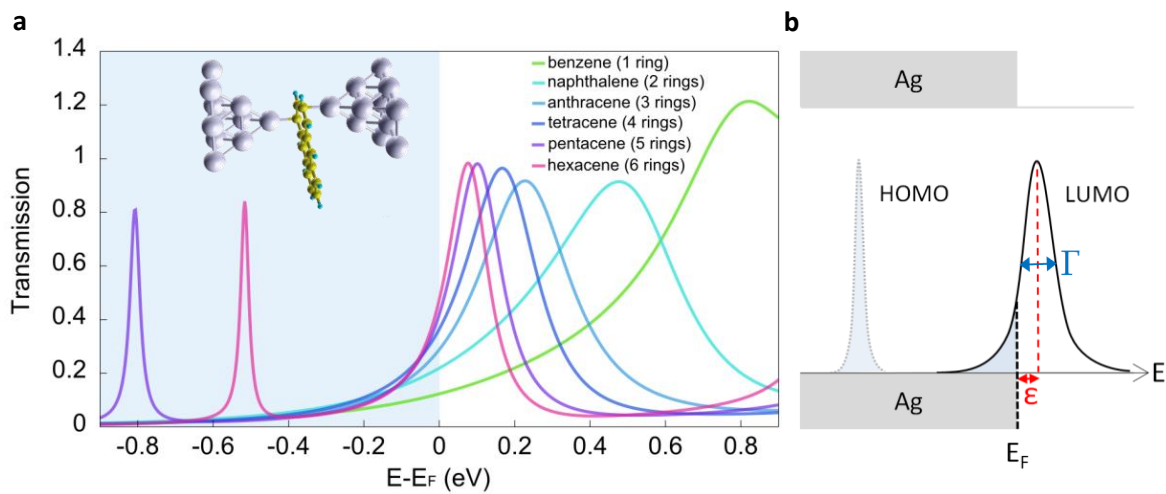

**Figure 3: Calculated transmission curves and the corresponding single Lorentzian model for the Ag/oligoacene junctions. a.** Calculated transmission curves for the Ag/oligoacene series. The conductance of each junction is determined by the transmission value at $E_F$ (E-$E_F$=0). The transmission peaks with a significant value at $E_F$ are LUMO-type peaks, centered at energies higher than $E_F$. **Inset:** An example for a compact junction configuration (Ag/anthracene), used in the calculations. **b.** Illustration of $\varepsilon$, the separation between $E_F$ and the center of the peak (red) and of $\Gamma$, the width of the peak (blue) in the framework of a single Lorentzian LUMO-type transmission peak that determines the conductance.

Despite the complexity of molecular junctions, an intuition about these two trends can be obtained from simple physical considerations, based on the length of the oligoacene molecules and their conjugated nature. In analogy to a "particle in a box" (Fig. 4a), the HOMO-LUMO gap of the oligoacenes, related to $\varepsilon$, is reduced with molecule length[23]. The latter can be expressed by N, the number of benzene rings comprising the molecule. In this context, $\varepsilon$ can be written as: $\varepsilon \propto (2n+1)/N^2$. Here, $n$ stands for the quantum number of the molecular π orbital that dominates the conductance and it is proportional to $N$, since the number of π electrons increases linearly with the number of conjugated rings, increasing $n$ accordingly. This results in (1) $\varepsilon = c_1/N + c_2/N^2$ ($c_1$, $c_2$ constants). When the molecules are attached to metal electrodes, the energies of the molecular levels change considerably. However, the general tendency of reduction in the HOMO-LUMO gap with length is preserved, implying a better alignment of these orbitals with $E_F$[5, 8, 11, 24], and an increase of conductance. In order to understand the onset of saturation that follows the conductance increase, it is important to consider another factor, the broadening $\Gamma$ of the LUMO peak, which results from the coupling of the LUMO to the extended states of the electrodes[1, 24]. Since the molecular π-orbitals are normalized over the whole molecule, the spatial weight of each orbital decreases with molecule length as ~1/N (Fig. 4b). As a result, their overlap with the valence metal orbitals decreases, yielding (2) $\Gamma=c_3/N$ ($c_3$ constant). The reduction in $\Gamma$ with molecule length acts to decrease the conductance, as opposed to the reduction in $\varepsilon$.

The resulting conductance trend is determined by the interplay between $\varepsilon$ and $\Gamma$, as evident from the expression for conductance in the single Lorentzian model: (3) $G=G_0/[(\varepsilon/\Gamma)^2+1]$. By introducing equations (1) and (2) this expression is reduced to: (4) $G=G_0/[(c_a+c_b/N)^2+1]$ ($c_a=c_1/c_3$, $c_b=c_2/c_3$). For short molecules the conductance increases with molecule length. However, for long enough molecules (large N) the term $c_b/N$ becomes negligible and the conductance saturates. A fit of equation (4) to the measured conductance trend is shown in the inset of Fig. 4c. The corresponding Lorentzian transmission peaks for $N=2,4,6$ are shown in Fig. 4c, demonstrating conductance increase (2→4) and the onset of conductance saturation (4→6).

As long as a single level model applies, the ratio $\varepsilon/\Gamma$ is the only dimensionless parameter. It indicates not only the conductance but also the charge-transfer between the electrodes and the molecular level. According to Friedel's sum rule, the conductance can be expressed uniquely through the charging of the level $q$, $G=G_0 \cdot \sin^2(\frac{\pi}{2} \cdot q)$ (see Supplementary Information). The relation shows that saturation of the conductance, which is a non-equilibrium quantity, results from the saturation of the charge, which is an equilibrium quantity. Thus conductance saturation at $G$~0.5-0.6 $G_0$ is associated with charge transfer of $q$~0.5-0.6 electrons to the molecule. In view of the above, the conductance saturation can be modeled even in situations where the assumptions derived from a "particle in a box" do not apply or when $\varepsilon$ is not proportional to the HOMO-LUMO gap.

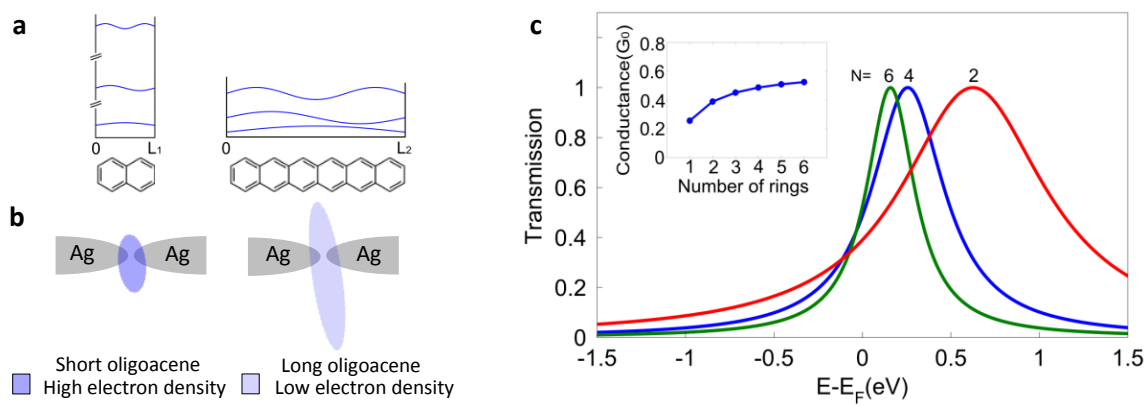

**Figure 4: A single level model explaining the conductance trend along the Ag/oligoacene series**. **a.** A "particle in a box" analogy for the energy level spacing of the oligoacenes. Increasing molecule length leads to a decrease in the energy level spacing. **b.** Schematic Illustration of the orbital spatial weight on the oligoacenes, demonstrated by the color intensity. Due to wave-function normalization, the spatial weight decreases with the length of the molecule, resulting in a lower overlap with the electrodes. **c.** A fit of equation (4) to the measured conductance trend along the Ag/oligoacene series (fitting parameters: $c_a=0.8$, $c_b=0.9$) and examples for the corresponding calculated Lorentzian peaks for $N=2,4,6$.

A remarkably different conductance behavior is observed when similar experiments are performed using Pt electrodes, which in contrast to Ag have prominent *d*-valence orbitals available for the conductance. As shown in Fig. 2 (upper purple triangles), the series of Pt\oligoacenes has a very high characteristic conductance of ~1 $G_0$, which is independent of the molecule length, namely conductance saturation along the entire series. This saturation, however, is of a different origin than the one observed for the Ag/oligoacene system. Fig. 5a shows typical conductance traces and histograms for bare Pt and Pt/oligoacene junctions. While the bare Pt junctions yield a conductance peak at ~1.5 $G_0$[25, 26], a peak at ~1 $G_0$ emerges upon introducing the molecules. Previous studies have found that the type of metal electrode in molecular junctions can change the conductance values[8, 9]. However, the general conductance trend was preserved. Here, the two sets of oligoacene-based molecular junctions (Fig. 2) show substantially different conductance trends.

To study the origin of conductance insensitivity to molecule length in the Pt/oligoacene junctions, we compare the transmission curves of Ag/pentacene and Pt/pentacene (Fig. 5b). Similar results were obtained for other oligoacenes. The transmission curve of Ag/pentacene (dotted purple) is composed of separate resonances, originating from the molecular levels. Therefore, the conductance is determined by the exact position and width of the relevant molecular level (LUMO) relative to $E_F$. On the other hand, the transmission curve of Pt/pentacene (light blue) reveals a band-like transmission. The very large broadening smears out molecular features and creates a broad plateau of high transmission, which is considerably less sensitive to changes in the molecular level alignment with respect to $E_F$. This leads to approximately constant conductance that does not follow the variations in the electronic structure of the oligoacenes along the series. The large broadening indicates strong hybridization between the valence orbitals of Pt to the conductive molecular orbitals. This observation is supported by former calculations that found a significant hybridization between the *d*-orbitals of Pt and the molecular *π*-orbitals in similar systems[27, 28]. In the case of Ag, due to insignificant contribution from *d*-orbitals around $E_F$, the transmission is mostly determined by the less efficient hybridization of the *s* valence orbital with the molecular *π*-orbitals and thus the junction preserves better the molecular character. Therefore, the different electronic nature of the metal electrodes results in significantly different hybridization, revealing two physical origins for conductance saturation.

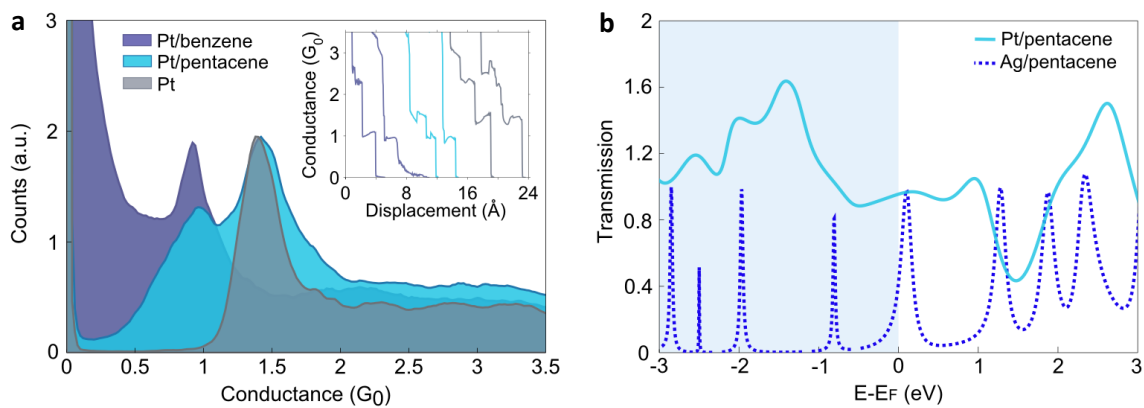

**Figure 5. Experimental and theoretial conductance characterization of the Pt/oligoacene molecular junctions a.** Conductance traces (inset) and conductance histograms recorded during pulling of Pt/benzene (purple) and Pt/pentacene (light-blue), compared to bare Pt (grey). Upon introducing oligoacenes to the Pt junction, a characteristic peak at ~1 $G_0$ appears, alongside or replacing the characteristic peak of the bare Pt at ~1.5 $G_0$. Each histogram is composed from at least 5,000 traces, recorded at a bias voltage of 200mv. **b.** Calculated transmission curve of Pt/pentacene (light blue), showing a broad, band-like plateau of high transmission, reflecting insensitivity of the conductance to the exact location of $E_F$. In comparison, the transmission curve of Ag/pentacene (dashed blue) is composed of resonances that originate from the molecular levels and the conductance is sensitive to the exact alignment of the LUMO resonance with respect to $E_F$.

To conclude, we study the conductance dependence on molecule length in a series of single molecule junctions at the high transmission regime. We find that the conductance can reach an upper limit, where it is independent of molecule length. By comparing two series of molecular junctions, both based on oligoacenes with increasing length, albeit with a different metal as electrodes, we have identified two fundamental mechanisms for the emergence of this conductance saturation. For Ag/oligoacene junctions, we find that the conductance increases with molecule length, followed by the onset of conductance saturation at a value determined by the interplay between level alignment and coupling strength. In contrast, Pt/oligoacene junctions are characterized by a band-like transmission, resulting in approximately constant conductance along the series of molecular junctions, regardless of the molecule length. Thus, by the right choice of metal electrodes, one can either obtain tunability of conductance, or alternatively achieve a robust metallic-like conductance across molecular junctions. Our findings provide insight into the conductance properties of metal-molecule interfaces near the full transmission limit, which is central for the realization of highly conductive metal-molecule interfaces.

**Methods**

Experimental:

The experiments are performed using a mechanical controllable break junction (MCBJ) setup at 4.2K. The sample is fabricated by attaching a notched Ag (99.997%, 0.1mm, Alfa Aesar) or Pt wire (99.99%, 0.1mm, Goodfellow) to a flexible substrate (1mm thick phosphor-bronze covered by 100µm insulating Kapton film). A three-point bending mechanism is used to bend the substrate in order to break the wire at the notch under cryogenic conditions and form an adjustable gap between two ultra-clean atomically sharp tips. A piezoelectric element (PI P-882 PICMA) is used to tune the bending of the substrate and control the distance between the electrodes with sub-angstrom resolution. The piezoelectric element is driven by a DAQ card (16 bit NI-PCI6221 or NI-PCI4461) connected to a high peak current piezo driver (Piezomechanik SVR 150/1). An ensemble of junctions with diverse structures is studied by repeatedly pushing the electrodes together to form a contact of ~50 $G_0$ and pulling the electrodes apart until full rapture, using the piezo element, at a rate of 20-40 Hz, and simultaneously measure the conductance. The junction is biased with a DC voltage provided by the DAQ card and divided by 10 to improve the signal to noise ratio. The resulting current is amplified by a current preamplifier (SR570) and recorded by the DAQ card at a sampling rate of 50-200 kHz. The oligoacenes (purity>99.9%) were purchased from Sigma-Aldrich, except for the hexacene, which was prepared according to a published procedure[17], where the final stage of synthesis was performed in-situ in the measurement vacuum chamber.

Calculations:

Density functional theory with the generalized gradient approximation to the exchange-correlation functional are performed[29]. The wave-functions are represented in a localized basis set, as implemented in the FHI-AIMS package[30]. We use the tier2 basis set, similar to the "double-zeta + polarization", common in quantum chemistry. Based on our previous findings, we performed closed shell calculations[18]. The electrodes are modeled as finite pyramidal clusters, cut from a face-centered crystal in the (111) direction. As a first step, a set of relaxed geometries is obtained by optimizing the positions of all molecular and both apex atoms. The electrodes used for the geometry optimization contain up to 11 Ag (Pt) atoms. In the second step, the geometry is fixed and transport calculations are performed. The transmission of Kohn-Sham electrons is calculated by non-equilibrium Green's function method for finite clusters[31] with the AITRANSS package[32]. In Ag-based junctions, the resonances are narrow (70meV for hexacene). The conductance is thus sensitive to the level alignment. The latter is determined by screening of the excess charge. In order to ensure proper screening by the electrode clusters, additional layers of Ag (Pt) atoms were added to both electrodes. With Ag electrodes, only for pyramids with 101 atoms (8 layers) or more, the resonance energy of the LUMO varies less than a few percent. With Pt electrodes, pyramids with 55 atoms (6 layers) are sufficient, because the conductance is less sensitive to shifts of the broad resonances (typical width is 1eV, see Fig. 5).

**Acknowledgements**

T.Y and O.T. thank L. Goffer, B. Pasmantirer, and K. L. Narasimhan for their valuable help in developing our measurement setups. O.T. thanks the Harold Perlman family for their support and acknowledges funding by the Israel Science Foundation, and the Minerva Foundation. R.K. and F.E. gratefully acknowledge the Steinbuch Centre for Computing (SCC) for providing computing time on the computer HC3 at Karlsruhe Institute of Technology (KIT). Part of the computational work was performed on the bwUniCluster resources funded by the Ministry of Science, Research and Arts and the Universities of the State of Baden-Wuerttemberg, Germany, within the framework programme bwHPC.

**Author contributions**

O.T and T.Y conceived the project and designed the experiments. T.Y performed the experiments with assistance from N.S and R.V. T.Y analyzed the data. R.K and F.E performed the calculations and participated together with T.Y and O.T in the overall analysis of the results. B.K and C.N synthesized the final precursor for hexacene. O.T and T.Y. wrote the paper and all co-authors commented on the manuscript.

**Additional information**

Supplementary information is available.

**References**

1.  Ratner, M.A. Introducing molecular electronics. *Materials Today* **5**, 20-27 (2002).
2.  Rocha, A.R. et al. Towards molecular spintronics. *Nat Mater* **4**, 335-339 (2005).
3.  van der Molen, S.J. & Liljeroth, P. Charge transport through molecular switches. *Journal of Physics: Condensed Matter* **22**, 133001 (2010).
4.  Pauly, F., Viljas, J. & Cuevas, J. Length-dependent conductance and thermopower in single-molecule junctions of dithiolated oligophenylene derivatives: A density functional study. *Physical Review B* **78**, 035315 (2008).
5.  Quek, S.Y., Choi, H.J., Louie, S.G. & Neaton, J.B. Length Dependence of Conductance in Aromatic Single-Molecule Junctions. *Nano Letters* **9**, 3949-3953 (2009).
6.  Livshits, G.I. et al. Long-range charge transport in single G-quadruplex DNA molecules. *Nat Nano* **9**, 1040-1046 (2014).
7.  Quinn, J.R., Foss, F.W., Venkataraman, L., Hybertsen, M.S. & Breslow, R. Single-molecule junction conductance through diaminoacenes. *Journal of the American Chemical Society* **129**, 6714-6715 (2007).
8.  Kim, B., Choi, S.H., Zhu, X.Y. & Frisbie, C.D. Molecular Tunnel Junctions Based on π-Conjugated Oligoacene Thiols and Dithiols between Ag, Au, and Pt Contacts: Effect of Surface Linking Group and Metal Work Function. *Journal of the American Chemical Society* **133**, 19864-19877 (2011).
9.  Kim, T., Vázquez, H., Hybertsen, M.S. & Venkataraman, L. Conductance of Molecular Junctions Formed with Silver Electrodes. *Nano Letters* **13**, 3358-3364 (2013).
10. Diez-Perez, I. et al. Controlling single-molecule conductance through lateral coupling of [pi] orbitals. *Nat Nano* **6**, 226-231 (2011).

11. Kaliginedi, V. et al. Correlations between molecular structure and single-junction conductance: A case study with oligo (phenylene-ethynylene)-type wires. *Journal of the American Chemical Society* **134**, 5262-5275 (2012).
12. He, J. et al. Electronic Decay Constant of Carotenoid Polyenes from Single-Molecule Measurements. *Journal of the American Chemical Society* **127**, 1384-1385 (2005).
13. Cheng, Z.L. et al. In situ formation of highly conducting covalent Au-C contacts for single-molecule junctions. *Nat Nano* **6**, 353-357 (2011).
14. Rascón-Ramos, H., Artés, J.M., Li, Y. & Hihath, J. Binding configurations and intramolecular strain in single-molecule devices. *Nat Mater* (2015).
15. Ferrer, J. & García-Suárez, V.M. Tuning the conductance of molecular junctions: Transparent versus tunneling regimes. *Physical Review B* **80**, 085426 (2009).
16. Venkataraman, L., Klare, J.E., Nuckolls, C., Hybertsen, M.S. & Steigerwald, M.L. Dependence of single-molecule junction conductance on molecular conformation. *Nature* **442**, 904-907 (2006).
17. Watanabe, M. et al. The synthesis, crystal structure and charge-transport properties of hexacene. *Nature chemistry* **4**, 574-578 (2012).
18. Korytár, R., Xenioti, D., Schmitteckert, P., Alouani, M. & Evers, F. Signature of the Dirac cone in the properties of linear oligoacenes. *Nat Commun* **5** (2014).
19. Muller, C.J., van Ruitenbeek, J.M. & de Jongh, L.J. Experimental observation of the transition from weak link to tunnel junction. *Physica C: Superconductivity* **191**, 485-504 (1992).
20. Pauly, F. et al. Molecular dynamics study of the thermopower of Ag, Au, and Pt nanocontacts. *Physical Review B* **84**, 195420 (2011).
21. Limot, L., Kröger, J., Berndt, R., Garcia-Lekue, A. & Hofer, W. Atom Transfer and Single-Adatom Contacts. *Physical Review Letters* **94**, 126102 (2005).
22. Untiedt, C. et al. Formation of a Metallic Contact: Jump to Contact Revisited. *Physical Review Letters* **98**, 206801 (2007).
23. Murov, S.L., Carmichael, I. & Hug, G.L. Handbook of photochemistry (CRC Press, 1993).
24. Moth-Poulsen, K. & Bjornholm, T. Molecular electronics with single molecules in solid-state devices. *Nat Nano* **4**, 551-556 (2009).
25. Nielsen, S. et al. Conductance of single-atom platinum contacts: Voltage dependence of the conductance histogram. *Physical Review B* **67**, 245411 (2003).
26. Krans, J. et al. One-atom point contacts. *Physical Review B* **48**, 14721-14724 (1993).
27. Yelin, T. et al. Atomically Wired Molecular Junctions: Connecting a Single Organic Molecule by Chains of Metal Atoms. *Nano letters* **13**, 1956-1961 (2013).
28. Ma, G. et al. Low-bias conductance of single benzene molecules contacted by direct Au–C and Pt–C bonds. *Nanotechnology* **21**, 495202 (2010).
29. Perdew, J.P., Burke, K. & Ernzerhof, M. Generalized Gradient Approximation Made Simple. *Physical Review Letters* **77**, 3865-3868 (1996).
30. Blum, V. et al. Ab initio molecular simulations with numeric atom-centered orbitals. *Computer Physics Communications* **180**, 2175-2196 (2009).
31. Arnold, A., Weigend, F. & Evers, F. Quantum chemistry calculations for molecules coupled to reservoirs: Formalism, implementation, and application to benzenedithiol. *The Journal of Chemical Physics* **126**, 174101 (2007).
32. Bagrets, A. Spin-Polarized Electron Transport Across Metal–Organic Molecules: A Density Functional Theory Approach. *Journal of Chemical Theory and Computation* **9**, 2801-2815 (2013).

**Conductance saturation in a series of highly transmitting molecular junctions**

**Supplementary Information**

T. Yelin[1], R. Korytár[2], N. Sukenik[1], R. Vardimon[1], B. Kumar[3], C. Nuckolls[3], F. Evers[2], O. Tal[1]

1. Department of Chemical Physics, Weizmann Institute of Science, Rehovot, Israel.
2. Institut für Theoretische Physik, Universität Regensburg, Regensburg, Germany.
3. Department of Chemistry, Columbia University, New York, United States.

**Friedel's sum rule for pedestrians**

We consider a molecule coupled to left and right metallic electrodes. We will demonstrate an explicit relation between the zero-bias conductance and the occupancy of the molecular level. We will assume that at low temperatures and zero bias, the electronic transport is governed by a single resonant orbital. Oligoacenes (starting from naphthalene) coupled to silver electrodes fall into this class, as we have demonstrated in the main text with the aid of first-principles calculations.

The transmission function T (ω) is given by

$$T(\omega) = \frac{\Gamma^2}{(\omega-\varepsilon)^2+\Gamma^2} \tag{S1}$$

where ε is the resonance center and 2Γ is the full resonance width at half maximum. The finite width of the resonance is induced by the electrodes and on writing equation (S1) we have assumed that the resonant state couples equally to both electrodes. The zero-bias conductance is proportional to the value of the transmission at the Fermi level

$$G = \frac{2e^2}{h}T(E_F) := G_0 T(E_F)$$

Without loss of generality, we set $E_F = 0$. In other words, ε is the resonance center with respect to the Fermi level. The spectral function A (ω) (density of states) is given by

$$A(\omega) = \frac{1}{\pi}\frac{\Gamma}{(\omega-\varepsilon)^2+\Gamma^2} \tag{S2}$$

The number of electrons in the resonant state is given by integration up to the Fermi level,

$$q = 2\int_{-\infty}^{0} A(\omega)d\omega = 1 - \frac{2}{\pi}arctan\left(\frac{\varepsilon}{\Gamma}\right) = \frac{2}{\pi}arccot\left(\frac{\varepsilon}{\Gamma}\right) \tag{S3}$$

Note the factor of two due to spin degeneracy. We introduce the phase shift

$$cotan(\eta) := \frac{\varepsilon}{\Gamma}, \eta = \frac{\pi}{2}q$$

which allows expressing the conductance as a function of the electron number only,

$$G = G_0 \frac{1}{1+\left(\frac{\varepsilon}{\Gamma}\right)^2} = G_0 \frac{1}{1+cotan^2(\eta)} = G_0 sin^2(\eta)$$

$$G = G_0 sin^2\left(\frac{\pi}{2}q\right) \tag{S4}$$

If the transport through the molecule in question is dominated by the LUMO, then $q$ represents the number of excess electrons in the LUMO in equilibrium. We note that the total charge transfer may contain components from other orbitals. The relation (S4) is known as Friedel's sum rule for historic reasons. In various forms, it has proven useful in the study of dilute alloys[1], adsorbed molecules[2] and transport through quantum dots[3].

We note here, that this relation holds also in the presence of electron-electron interactions, provided that certain Fermi-liquid identities apply[1,4]. In simpler terms, interactions in the Fermi-liquid state preserve the form of the spectral function and transmission resonance, equations S1 and S2, hence the same relation applies. This is the case of normal metallic electrodes at sufficiently low temperatures except for superconducting environments, ferromagnetic environments or other cases with interaction-induced broken symmetries. If the molecule develops a magnetic moment, Friedel's sum rule applies in the low-temperature fully-screened Kondo regime.

**References**

1   Hewson, A. C. *The Kondo problem to heavy fermions* (Cambridge university press, 1997).

2   Desjonqueres, M.-C. *Concepts in Surface Physics: 2ème édition*. Vol. 30 (Springer Science & Business Media, 1996).

3   Pustilnik, M. & Glazman, L. Kondo effect in quantum dots. *Journal of Physics: Condensed Matter* **16**, R513 (2004).

4   Langer, J. S. & Ambegaokar, V. Friedel Sum Rule for a System of Interacting Electrons. *Physical Review* **121**, 1090-1092 (1961).